\documentclass[aps,prl,onecolumn,superscriptaddress]{revtex4}
\usepackage{graphicx}
\usepackage{dcolumn}
\usepackage{bm}
\usepackage{url}
\usepackage{amssymb, amsmath}

\begin{document}

\title{STDP-driven networks and the \emph{C. elegans} neuronal network}

\author{Quansheng Ren}
\affiliation{Max Planck Institute for Mathematics in the Sciences,
Inselstr. 22, D-04103 Leipzig, Germany}
\author{Kiran M. Kolwankar}
\affiliation{Max Planck Institute for Mathematics in the Sciences,
Inselstr. 22, D-04103 Leipzig, Germany} \affiliation{Department of
Physics, Ramniranjan Jhunjhunwala College, Ghatkopar (W), Mumbai 400
086, India}
\author{Areejit Samal}
\affiliation{Max Planck Institute for Mathematics in the Sciences,
Inselstr. 22, D-04103 Leipzig, Germany} \affiliation{Laboratoire de
Physique Th\'eorique et Mod\`eles Statistiques, CNRS and Univ
Paris-Sud, UMR 8626, F-91405 Orsay, France}
\author{J\"urgen Jost}
\affiliation{Max Planck Institute for Mathematics in the Sciences,
Inselstr. 22, D-04103 Leipzig, Germany} \affiliation{The Santa Fe
Institute, 1399 Hyde Park Road, Santa Fe, New Mexico 87501, USA}
\email{jost@mis.mpg.de}

\begin{abstract}
We study the dynamics of the structure of a formal neural network
wherein the strengths of the synapses are governed by
spike-timing-dependent plasticity (STDP). For properly chosen input
signals, there exists a steady state with a residual network. We
compare the motif profile of such a network with that of a real
neural network of \emph{C. elegans} and identify robust qualitative
similarities. In particular, our extensive numerical simulations
show that this STDP-driven resulting network is robust under
variations of the model  parameters.
\end{abstract}
\maketitle


\section{\label{sec:I}Introduction}

In any theoretical study on neuronal networks, the formal structure
of the network as a directed graph is an important ingredient which,
in reality, can be rather complicated. In the last decade,
statistical concepts and tools have been developed for analyzing the
large scale properties of complex
networks~\cite{complexnet01,complexnet02,complexnet03,complexnet04,complexnet05}.
On this theoretical basis, it has been found that neural networks
can exhibit scale free and small world
properties~\cite{Egu,brain_net}. For a more refined analysis and the
identification of deeper properties that may or may not distinguish
neuronal networks from other classes of biological or non-biological
networks, it
 is necessary to identify those  factors that determine the structure of  neuronal networks.
Naively, one might think that the structure of a neuronal network is
determined genetically. But for animals with larger brains, this
would require an enormous amount of genetically encoded information.
In other words,  at best the connections of a few important axons
could be genetically encoded. Moreover, several  experimental
findings we decribe below suggest a lack of any hard-wired programme
of axon guidance. Another possible factor could be geometric
constraint since the network is embedded in a small three
dimensional volume. This is likely to affect only the long range
connections and not the local structure, and in any case, this
constraint seems to be too unspecific. Therefore, we should expect
that some self-organization process yields the connectivity
structure of neuronal networks. As most biological self-organization
processes are triggered by external factors or signals, we should
also look here for sources of external influences. Since a neuronal
network processes sensory inputs, it should thereby adapt itself to
its experiences. Hence, we are naturally led to consider learning as
a key factor guiding the self-organization of a neuronal network.
The standard Hebbian paradigm tells us that learning is represented
by modifications of
 the strengths of the synapses between neurons. In particular,
 learning is local in the sense that it depends on correlations
 between the activities of synaptically connected neurons. In more
 biological detail, we have the learning scheme of \emph{spike timing
   dependent synaptic plasticity} first discovered in \cite{Markram},
 abbreviated as STDP. This learning rule says that a synapse is
 strengthened when the presynaptic neuron fires shortly before the
 postsynaptic one, and that it is weakened instead when this temporal
 order is reversed. This learning rule has received a lot of attention
 in the neurobiological literature. In \cite{J}, it has been formally
 analyzed how this learning rule, being based on activity
 correlations, in turn  shapes these correlations.  Again, this is a local rule, but it is then
 natural to expect that the global statistical properties of neuronal
 networks result from the iterated application of this local rule at
 all the active synapses of the network. In particular, some synapses
 could possibly become so weak that they will get eliminated entirely.

 The initial explorations
to test this line of thinking have been encouraging~\cite{JK,SK}.
In~\cite{JK}, in order to separate the abstract features of this
learning rule from the details of its neurobiological
implementation, we have considered a simple model of coupled chaotic
maps wherein the coupling stengths changed according to this
learning rule. Starting from a globally coupled network, we obtained
a stationary network with a broad degree distribution in accordance
with the experimental findings for real neural networks.
In~\cite{SK}, similar conclusions were arrived at using a continuous
Fitzugh-Nagumo model.

These developments suggest that the learning dynamics may be a
relevant factor in determining the network structure. A closer
comparison is needed to confirm this conjecture. To this effect we
carry out simulations with realistic models of neural dynamics and
compare the resultant network with a real one, that is, the neuronal
network of \emph{C. elegans}. The neuronal network of \emph{C.
elegans} has been studied in detail~\cite{White,Varshney:Celegans}.
In fact, this is the only real neuronal network where such detail is
currently available, and, unfortunately, we therefore are not able
to use other experimentally determined networks for comparison. In
any case, in learning, there are various kinds of plasticities and
different kinds of connections, like, chemical synapses, gap
junctions etc. Here we work with realistic neuron models, the STDP
learning rule and properly chosen input signals. We show that
applying this scheme  formally leads to a network which is similar
to the network in \emph{C. elegans} in certain aspects.

It is known that the brain has a very dense population of synaptic
connection just after the birth and most of these connections are
pruned in the course of time~\cite{Bis}. This type of pruning takes
place even in \emph{C. elegans} where the size of the network is
very small~\cite{Prun_CE}. It has also been shown that the perturbed
sensory activity or the mutations that alter the calcium channels or
membrane potential affect the axon outgrowth~\cite{Sensory}. This is
reflected by the irreversible deletion of synapses whose strength
falls below a certain threshold.

The plan of the paper is as follows. In section~\ref{sec:II}, we
decribe the neuron models used, the STDP learning rule and also the
tools used to analyse the network. The present status of our
knowledge of \emph{C. elegans} network is recalled in
section~\ref{sec:III}. This section also includes a more detailed
analysis of \emph{C. elegans}' neuronal network. The main results of
the paper are presented in  section~\ref{sec:IV}; in particular, we
describe the influence of the input and of different parameters on
the final results. The paper ends with a discussion in
section~\ref{sec:V}.

\section{\label{sec:II}Methods}

\subsection{\label{sec:IIA}Neuron models}
Networks of neurons were modeled using the NEST Simulation Tool
\cite{Gewaltig:NEST}. To show the generality of the results, two
models for neurons are utilized, i.e. the Leaky Integrate-and-Fire
(LIF) model and the Hodgkin-Huxley (HH) model.

The membrane potential $V_j$ of the conductance based LIF neuron
with index $j$ is governed by
\begin{equation}
    C_m\frac{dV_j}{dt}=g_L(V_{rest}-V_j)+g_j(t)(E_{ex}-V_j)\;
    \label{eq:LIF},
\end{equation}
where $C_m=200$pF is the membrane capacitance,  $g_L=10$nS is the
leak conductance which is equivalent to $R_m=100\mbox{M}\Omega$
where $R_m$ is the membrane resistance, $V_{rest}=-70$mV is the
resting potential (leak reversal potential), $E_{ex}=0$mV is the
excitatory reversal potential. In our simulation, we did not
consider  inhibitory synapses as they are extremely rare between
interneurons in the \emph{C. elegans} neuronal network (see
Section~\ref{sec:III}). When the membrane potential reaches the
threshold value $V_{th}=-54$mV, the neuron emits an action
potential, and the depolarization is reset to the reset potential
$V_{reset}=-60$mV after a refractory period $\tau_{ref}=1$ms during
which the potential is insensitive to stimulation. The parameters
given above are the same as in \cite{Song:stdp}.

The dynamical equation for the Traub modified conductance based
Hodgkin-Huxley model neuron is
\begin{equation}
  C_m\frac{dV_j}{dt}=g_L(E_L-V_j)+g_{Na}m^3h(E_{Na}-V_j)+g_Kn^4(E_K-V_j)+g_j(t)(E_{ex}-V_j),
\label{eq:HH}
\end{equation}
where $C_m=100$pF. $E_{ex}=0$mV is the excitatory reversal
potential. The maximal conductances and reversal potentials of the
sodium and potassium ion channels and the leak channel used in the
model are $g_{Na}=1.0\mbox{mS/mm}^2$, $g_K=2.0\mbox{mS/mm}^2$,
$g_L=0.001\mbox{mS/mm}^2$, $E_{Na}=48$mV, $E_K=-82$mV, and
$E_L=-67$mV respectively. The gating variables $X={m,h,n}$ satisfy
the following equation:
\begin{equation}
  \frac{dX}{dt}=\alpha_X(V_j)(1-X)-\beta_X(V_j)X,
\end{equation}
where $\alpha_X$ and $\beta_X$ are given by
\begin{eqnarray*}
\alpha_m=\frac{0.32(V+54)}{1-\exp(-0.25(V+54))} & & \beta_m=\frac{0.28(V+27)}{\exp(0.2(V+27))-1}\\
\alpha_h=0.128\exp(-(V+50)/18) & & \beta_h=\frac{4}{1+\exp(-0.2(V+27))}\\
\alpha_n=\frac{0.032(V+52)}{1-\exp(-0.2(V+52))} & &
\beta_n=0.5\exp(-(V+57)/40).
\end{eqnarray*}
These parameters  are taken from \cite{Traub:HHmodel}.

The synaptic conductance $g_j(t)$ in Eq.~(\ref{eq:LIF}) and
(\ref{eq:HH}) is determined by
\begin{equation}
  g_j(t)=g_{m}\sum_{j=1}^{N}w_{ij}(t)\sum_{k}f(t-t_j^k)\;
  \label{eq:LIF2},
\end{equation}
where N is the number of neurons, $g_{m}$ is the maximum value of
the synaptic conductance, $w_{ij}$ is the weight of the synaptic
connection from the $i$th neuron to the $j$th neuron, $t_j^k$ is the
timing of the $k$th spike of the $j$th neuron. Here, we used an
$\alpha$-function \cite{Brunel:alphafunc} $f(x)$ with latency
(transmission delay) $\tau_d$ and synaptic time constant
$\tau_{ex}=2$ms:
\begin{equation}
  f(t)=
  \begin{cases}
    \frac{t-\tau_d}{\tau_{ex}^2}\exp(-\frac{t-\tau_d}{\tau_{ex}}) & \mbox{if $t>\tau_d$} \\
    0 & \mbox{otherwise}.
  \end{cases}
\end{equation}

\subsection{\label{sec:IIB}STDP}
STDP is a form of experimentally observed (\cite{Markram}) long-term
synaptic plasticity, where synapses are modified by repeated
pairings of pre- and postsynaptic action potentials, while the sign
and the degree of the modification depend on their relative timing.

In our study, the weight of the synaptic connection $w_{ij}$ is
modified by the STDP rule. The amount of modification is determined
based on the temporal difference $\Delta t$ between the occurrence
of the postsynaptic action potential and the arrival of the
presynaptic action potential,
\begin{equation}\label{eq:delta-t}
  \Delta t=t_j-(t_i+\tau_d),
\end{equation}
where $t_j$ is the spike time of the postsynaptic neuron $j$,
$\tau_d$ is the delay time of the spike transmission from neuron $i$
to neuron $j$, and $t_i$ is the spike time of the presynaptic neuron
$i$. The weight modification $\Delta w_{ij}$ is described by the
following equations:
\begin{equation}\label{eq:STDP}
  \Delta w_{ij}(\Delta t)=
  \begin{cases}
    \lambda\exp(-|\Delta t|/\tau_+) & \mbox{if $\Delta t\geq\tau_d$} \\
    -\lambda\alpha\exp(-|\Delta t|/\tau_-) & \mbox{if $\Delta t<\tau_d$},
  \end{cases}
\end{equation}
where $\lambda=0.0001$ is the learning rate. We constrain $w_{ij}$
within the range $[0,1]$, which ensures that the peak synaptic
conductance $g_mw_{ij}$ is always positive and cannot exceed the
maximum value $g_m$. In Eq. \ref{eq:STDP}, $\alpha$ introduces a
possible asymmetry between the scale of potentiation and depression.
The time constants $\tau_+$ and $\tau_-$ control the width of the
time window. As argued in~\cite{Song:stdp}, in order to get a stable
competitive synaptic modification, which means that uncorrelated
pre- and postsynaptic spikes produce an overall weakening of
synapses, the integral of $\Delta w_{ij}$ should be negative. A
negative integral requires $(\alpha\tau_-/\tau_+)>1.0$.

\subsection{\label{sec:IIC}Network Motifs}
To test whether STDP plays a crucial role in the evolution of a real
neuronal network, it is important to compare the local structure
between real networks and the network we obtain as a result of our
simulations. We choose to look at the occurrence of different
network motifs. Network motifs are patterns (sub-graphs) that recur
within the network much more often than expected at random
\cite{Milo:motif}. The characterisation of networks using network
motifs has become very common owing to the fact that different
subnetworks are thought to carry out different functions in the
network and the abundance of certain subnetworks can decide the
overall character of the network. In order to determine the relative
occurrence of motifs one needs to generate random versions of the
network and count the number of motifs. The question of the choice
of the null model and the method used to generate random networks is
important. We use the approach and the software mfinder  developed
by Alon and co-workers \cite{Milo:motif, Milo:superfamily}. First,
numbers of different three-node subgraphs in a given network are
found. Then, we compare the network to an ensemble of randomized
networks, whose number is 1000 in this study. Randomization is
performed by rewiring connections in such a way that the number of
incoming edges, outgoing edges and mutual edges of each node are
preserved. For each subgraph $i$, the statistical significance is
defined by its Z-score:
\begin{equation}\label{eq:Zscore}
  z_i=\frac{N_i^{real}-\langle N_i^{rand} \rangle}{std(N_i^{rand})},
\end{equation}
where $N_i^{real}$ is the frequency of subgraph $i$ appearing in the
real network, $\langle N_i^{rand} \rangle$ and $std(N_i^{rand})$ are
the mean and standard deviation of subgraph $i$'s occurrences in the
ensemble of random networks. If $z_i>0$, subgraph $i$ is
over-represented and is designated as a motif, while if $z_i<0$, it
is under-represented and is designated as a anti-motif. The
significance profile (SP) of a  network is the vector of Z-scores
normalized to length 1:
\begin{equation}\label{eq:SP}
    SP_i=\frac{z_i}{\sqrt{\sum_i^{13}z_i^2}}.
\end{equation}
It shows the relative significance of subgraphs and is important for
comparison of networks of different sizes and degree sequences. The
software mfinder \cite{mfinder} has integrated all the above
algorithms.

To construct random ensembles, one often uses Occam's razor, i.e. no
outcome compatible with the null hypothesis should be preferred and
all such outcomes are equally likely \cite{Foster:likelihood}. The
approach mentioned above uses the random ensemble with fixed degree
sequence and fixed number of 2-node subgraphs for each node as the
relevant null model. We should point out here that a more conceptual
approach  \cite{Kahle,Olbrich} leads to somewhat different null
hypotheses, as will be briefly discussed in Section~\ref{sec:V}. In
particular, the SPs of different motifs are not independent. That
is, a change in some motif can also affect the count of another
motif and hence its SP. In spite of these and many other
complications,  Milo \mbox{et al.} \cite{Milo:superfamily} have
discovered four superfamilies with distinct motif diagrams. The
second superfamily included signal-transduction networks,
developmental transcription networks of multi-cellular organisms and
the \emph{C. elegans} neuronal network. The existence of such
``universality classes" of different natural as well as artificial
networks makes this approach  of characterising complex networks,
which are otherwise rather untractable, very promising. Several
explanations~\cite{Milo:superfamily, Mikhailov,KB} have been
proposed to explain the observed convergence of SPs of these
networks. In particular, robustness against node or link failure has
been invoked a global optimization criterion leading to such
particular motif distributions as in the superfamilies. Our study
provides an alternative explanation in terms of a local
self-organization rule, i.e. the common SP of the second superfamily
may come from an adaptive mechanism in cooperation with the complex
structure of correlations between input signals.

\section{\label{sec:III}C. elegans neuronal network}
In this section we discuss the neuronal network of \emph{C.
elegans}. It is a small sensory transduction neuronal network that
consists of sensory neurons, interneurons and motor neurons. In this
study, the most recently published wiring diagram of \emph{C.
elegans}~\cite{Varshney:Celegans} is used. The somatic neuronal
network contains 279 neurons which are connected by 2194 directed
connections implemented by one or more chemical synapses, and 514
gap junction connections consisting of one or more electrical
junctions. We point out that, in contrast to vertebrates, there are
no individual variations among different members of the species
\emph{C. elegans} here. As STDP can only operate in chemical
synapses, we restrict our attention to the chemical synapse network.
However, one should note that the bidirectional gap connections
could influence the dynamics of neurons, which, in turn, could
influence the network topology further by plasticity mechanisms like
STDP.

\begin{figure}
\centering
\includegraphics[scale=.4]{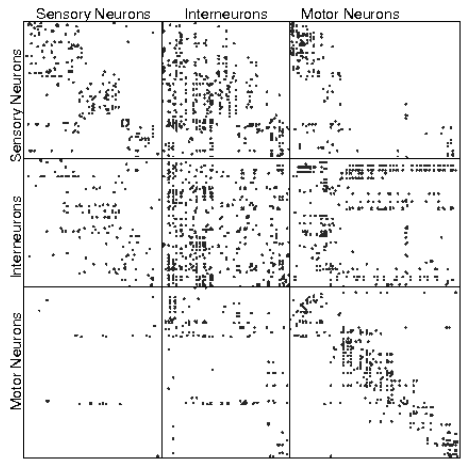}
\caption{\label{fig:CEmatrices} Adjacency matrix of the chemical
synapse network with neurons grouped by category (sensory neurons,
interneurons, and motor neurons). Within each category, neurons are
in anteroposterior order. The figure is reconstructed from the data
contained in~\cite{Varshney:Celegans}.}
\end{figure}

\begin{figure}
\centering
\includegraphics[width=14cm]{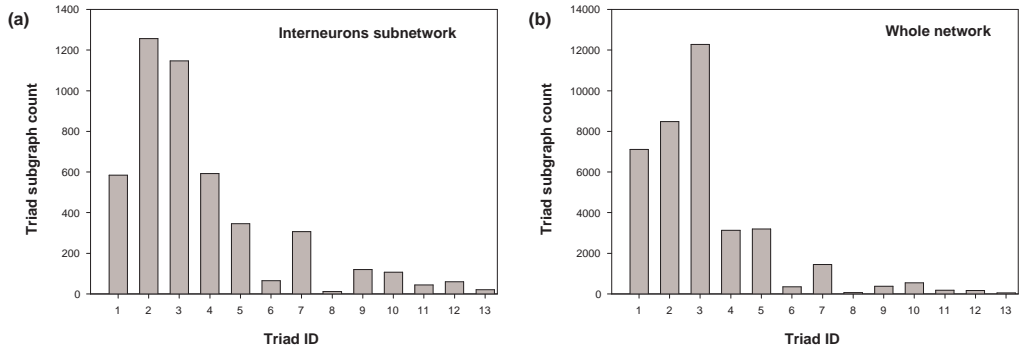}
\caption{\label{fig:CEfrequency} Triad subgraph frequency spectrum
of (a) the subnetwork of interneurons and (b) the whole network in
\emph{C. elegans}. The structure corresponding to each triad ID is
shown in fig.~\ref{fig:Triad}.}
\end{figure}

From fig.~\ref{fig:CEmatrices}, we see that the connections both
from sensory neurons to interneurons or from interneurons to motor
neurons are more numerous than the connections from interneurons to
sensory neurons or from motor neurons to interneurons respectively.
On the other hand, there are rare connections between sensory
neurons and motor neurons. External signals first arrive at sensory
neurons, then propagate through some interneurons and finally reach
a subset of motor neurons to generate a stimulus-induced response.
Both sensory neurons and motor neurons have preferred connections to
neighboring neurons and can be divided into several clusters, which
may correspond to different functions. However, the pool of
interneurons is not organized into clusters or layers. Each signal
passes through a few interneurons before it reaches motor neurons.
We want to study the network structure generated by the STDP-driven
pruning process from a pool of homogeneous neurons. From this point
of view, the pool of interneurons in \emph{C. elegans} is more
related to our case than sensory neurons and motor neurons. There
are 82 interneurons and 479 connections between them, among which
there are 122 mutual links and 357 uni-directional links. Moreover,
there is only 1 inhibitory synapse between interneurons according to
the latest data and a rough approximation that GABAergic synapses
are inhibitory \cite{Varshney:Celegans}. Fig.~\ref{fig:CEfrequency}
shows the triad subgraph frequency spectrum of the subnetwork of
interneurons and the whole neuronal network in \emph{C. elegans}.

The SP of the \emph{C. elegans} neuronal network has previously been
shown in~\cite{Milo:superfamily}. Some motif analysis of the latest
wiring diagram has also been presented in \cite{Varshney:Celegans}.
However, the local structure may be quite different for different
functional networks. Here we calculate SPs for different categories
of neurons in \emph{C. elegans} separately. Fig.~\ref{fig:CEsp}
shows the SPs of different subnetworks with only sensor neurons,
only motor neurons, only interneurons, the combination of sensor
neurons and interneurons, the combination of motor neurons and
interneurons, and the whole neuronal network respectively. Comparing
these SPs, we found that the SP characteristic of \emph{C. elegans}
neural system reported previously~\cite{Milo:superfamily,
Varshney:Celegans} mainly comes from the connectivity structure of
interneurons. The SP of interneurons subnetwork shows triads 7, 9
and 10 as motifs, triads 1, 2, 4, 5 and 6 as anti-motifs, and there
is less bias against cascades (triad 3). In this paper we focus on
the question whether these observed motifs could be generated via
self-organization through STDP.

\section{\label{sec:IV}Evolution of neural networks with STDP}

It is known that the density of synapses in the human frontal cortex
continues to increase during infancy and remains at a very dense
level. After a short stable period, synapses begin to be constantly
removed, yielding a decrease in synaptic density. This pruning
process continues until puberty, when synaptic density achieves
adult levels \cite{Huttenlocher:dense}. As such a pruning of
synapses that are in some sense superfluous may be a rather
universal process, we study the local structure of a network
obtained by an STDP-driven pruning process, and compare it with
\emph{C. elegans}.

\subsection{\label{sec:IVA} Basic phenomena}
To simulate the STDP-driven pruning process, we start the
simulations with an all-to-all connected network. The neurons are
stimulated by different periodic patterns repeatedly with period
$T_{pattern}$. All the patterns are truncated from Poisson spike
trains with the same average rate $f_{Poisson}=50$Hz. This average
firing rate corresponds to a 20ms spike interval which is consistent
with the width of STDP time window. The connections between neurons
are excitatory STDP synapses. Because GABAergic synapses are
extremely rare between interneurons of \emph{C. elegans}, we do not
consider inhibitory synapses here.

\begin{figure}
\centering
\includegraphics[width=12cm]{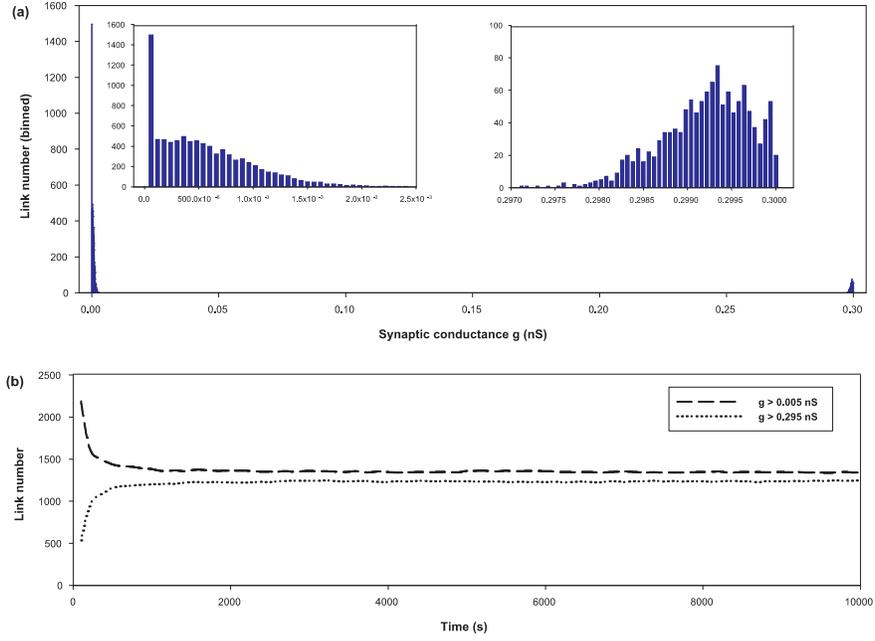}
\caption{\label{fig:STDP_A} (color online) (a) The final
distribution of peak synaptic conductances $g$. (b) The variation of
the link numbers of synapses whose peak conductance $g$ is above
0.005nS or 0.295nS. The parameter values are the same as in the
``basic configuration" described in text.}
\end{figure}

Because of the asymmetry between the scale of potentiation and
depression, most synapses are weakened during the learning process.
The peak synaptic conductances approach a bimodal distribution. Then
we filter the adjacency matrix by a small threshold: If the weight
of a synapse is less than the threshold, we consider it as a pruned
synapse, i.e. there is no connections, otherwise we consider it as a
winning synapse. At last, an unweighted adjacency matrix is
obtained. We analyze the local structure of the network, and compare
it with \emph{C. elegans}. To simulate the long term development of
neural systems, we used a small learning rate ($\lambda=0.0001$) and
simulated for more than $10^7$ms. All the peak synaptic conductances
and the potentials of neurons are initialized with a random uniform
distribution. After development, most of the peak synaptic
conductances are pushed toward zero or $g_m$ (cf.
fig.~\ref{fig:STDP_A}a). We set the threshold to $g=0.005$nS, under
which the synapses are seen as pruned. Fig.~\ref{fig:STDP_A}b shows
the variation in the number of links, i.e. the number of the
synapses whose peak conductance is above the threshold. We see that
the number of links decreases rapidly before $10^6ms$, and toward
the end the distribution of peak synaptic conductances remains
almost constant except for tiny fluctuations, i.e. a steady-state
condition is achieved. We then analyze the occurrence of triad
motifs in the resulting steady-state network.

Following the preceding approach, we study four cases with different
configurations. The first case is called the  ``basic
configuration", and later cases are  variations of this one. For
this ``basic configuration", we simulate a small network with
$N=100$ LIF neurons, which is of a similar size as the subnetwork of
somatic interneurons in \emph{C. elegans}. We used an asymmetric
time window $\tau_+=16.8$ms and $\tau_-=33.7$ms in STDP rule, which
provides a reasonable approximation of the synaptic modification
observed in actual experiments \cite{Poo:stdp}. $\alpha=0.525$ is
adopted, together with the asymmetric time window, providing a ratio
$A_-\tau_-/A_+\tau_+=1.05$ which is the same as in \cite{Song:stdp}.
Other parameters are set as follows: the synaptic delay
$\tau_d=10$ms, the maximum peak synaptic conductance $g_m=0.3$nS,
and the period of input patterns $T_{pattern}=2$s. Based on the
``basic configuration", we study three variations: ``Symmetric
configuration", where the asymmetric time window is replaced with a
symmetric one ($\tau_+=\tau_-=20.0$ms), and $\alpha=1.05$ to
preserve the ratio $A_-\tau_-/A_+\tau_+=1.05$; ``HH model
configuration", where LIF model is replaced by the HH model; ``Large
network configuration", where the network size is enlarged to 200
neurons, and $g_m=0.2$nS in this case. For every configuration, we
simulate 10 times with different input patterns, different initial
potential of neurons and different initial peak synaptic
conductances. We calculate the SP of every simulation and compute
the average of 10 SPs.

Fig.~\ref{fig:STDP_B} shows the SPs of the four configurations and
compares them with the \emph{C. elegans} neuronal network. All the
four STDP-driven evolved network have quite similar SPs. All of them
show triads 7, 9, and 10 as motifs, and triads 1, 2, 4, 5, as
anti-motifs, as in the \emph{C. elegans} neuronal network which
belongs to the second superfamily reported in
\cite{Milo:superfamily}. This phenomenon does not depend on the
neuron model, the symmetry of the time window or the network size,
and so must reflect some intrinsic characteristic of STDP. Recently,
several interesting functions and dynamics have been found to be
associated with the three motifs mentioned above \cite{Chunguang}.
For example, the feedforward loop (FFL, triad 7) has been shown to
perform signal-processing tasks such as acceleration and delay of
response, and the mixed-feedforward-feedback loops (MFFL1: triad 9,
and MFFL2: triad 10), where two-node feedbacks that regulate or are
regulated by a third node, have been shown to perform long- and
short-term memory. These functions are important for almost all
neural computation and cognition tasks, which may give an
explanation for the redundance of these motifs in the \emph{C.
elegans} neuronal network. Our results show that STDP may be a
potential mechanism which could develop these important motifs.

The SP curves of STDP-driven evolved networks are quite similar to
the \emph{C. elegans} neuronal network, especially the subnetwork of
interneurons, but not the subnetwork of sensor neurons or moter
neurons (see fig.~\ref{fig:CEsp}). This could mean that though STDP
determines the structure of the network of interneurons, possibly
other factors are important in the case of sensory neurons.
Nevertheless, there are small differences between our evolved
networks and the subnetwork of interneurons in \emph{C. elegans}. As
the SP is the vector of Z-scores normalized to length 1, here we
only need to consider the relative relations between triads as well
as the zero axis. In STDP-driven evolved networks, triads 1, 2 and 8
have relatively lower negative SP, while triads 3 and 7 have
relatively higher positive SP than the ones in the \emph{C. elegans}
neuronal network. STDP tends to form feedforward structures
\cite{Masuda:STDP_FF, Takahashi:STDP_FF} and reflect the causal
relations between neurons, which could lead to  more representations
of cascades (triad 3) and FFL (triad 7), and less representations of
cycles (triad 8). On the other hand, there are hundreds of gap
connections and other properties in the \emph{C. elegans} neuronal
network, which we neglect here. The gap connections could certainly
influence the dynamics of neurons, while the latter would influence
the network structure if there is a certain plasticity mechanism
like STDP.

To study the similarity further, we calculated the triad frequency
spectra of STDP-driven evolved networks and compare them with the
one of the \emph{C. elegans}  network. From fig.~\ref{fig:STDP_C} we
see that the STDP-driven evolved network of the ``basic
configuration" develops a similar triad frequency spectra as that of
\emph{C. elegans}.

\subsection{\label{sec:IVB} The role of neuronal inputs.}

It appears that the inputs received by a neuron play an important
role  in determining the final network structure. This is already
borne out in the experiments~\cite{Sensory}. We investigate this
point in our simulations. Besides periodic patterns obtained from
Poisson spike trains, we also stimulate neurons with other spike
trains, such as stochastic Poisson spike trains and periodic regular
patterns. In these cases, we either do not obtain a steady
distribution of peak synaptic conductances or else find a similar
SP. We identify  a factor that is crucial for the similarities of
the local structure between STDP-driven evolved networks and the
\emph{C. elegans} neuronal network: The complexity of correlations
between neuronal inputs.

There seem to be two ways to achieve a constant distribution of peak
synaptic conductances through STDP: One is to use constant temporal
correlations between inputs. The other is through the repeated input
of the same temporal sequences \cite{Rabinovich}. To investigate the
former case, we utilize a simple method to generate sufficiently
complex  temporally correlated inputs: A non-periodic Poisson spike
train, whose average firing rate is 50Hz, is randomly delayed for
100 times to generate the inputs of 100 neurons. The random delay
time $T_d$ follows a uniform distribution between [1ms, T]. We study
two cases: T=20ms and T=200ms. Considering the width of time window
in STDP and average spike interval (both $\approx$20ms), the case of
T=20ms corresponds to a high degree and a simple relation structure
of temporal correlations between neuronal inputs, while the case of
T=200ms corresponds to a more complex relation structure. From
fig.~\ref{fig:CORR}a we see that the case of T=200ms generates an SP
 quite similar to that of the interneurons in \emph{C.
elegans}, and the only evident difference is in Triad 12. However,
in the case of T=20ms we do not obtain such a result. This not only
confirms that STDP may play a fundamental role in the formation of
the local structure of neuronal networks, but also prompts us to pay
attention to the complexity of correlations between neuronal inputs.

On the other hand, this simple scheme of temporally correlated
inputs did not generate a  triad frequency spectrum similar to that
of \emph{C. elegans} (fig.~\ref{fig:CORR}b and c). In the case of
T=200ms, it generates  more triads belonging to a feedforward
structure, i.e. cascades (triad 3), FFL (triad 7) and MFFL (triad 9
and 10), but could not generate any subgraph of triads 6, 8 and 11.
This may be because the correlation scheme we utilize here is not
complex enough as compared to real cases. In \emph{C. elegans},
because of the stochastic characteristic in the environment, common
stimuli among adjacent sensory neurons and the sensory network
structure, inputs of interneurons could have very complex temporal
correlations. It seems difficult  to simulate the same temporal
correlated neuronal inputs as in \emph{C. elegans}.

However, the repeating of temporal sequences plays an alternative
role in generating complicated correlated neuronal inputs. The
finite size of sequences brings out non-trival temporal
correlations, which could be acumulated by repeating input. The
complexity of the non-trival correlations is determined by the input
sequences. Partial information of sequences is transformed into a
network of stronger or weaker synapses among the neurons. The
evolved local network structure reflects certain intrinsic
characteristic of input sequences. The temporal sequences we used is
truncated from a Poisson process, which provides an extremely useful
approximation of stochastic neuronal firing. It could provide enough
complexity in the structure of correlations between inputs, which
gives rise to a similar triad frequency spectrum as that of \emph{C.
elegans} (fig.~\ref{fig:STDP_C}). Otherwise, if regular sequences or
simply correlated inputs are used, we do not get similar results.

As the characteristic of correlations comes from a finite-size
effect and the characteristic of input sequences, it should not
depend on other factors, such as the period of repeating. To verify
this, we study two more cases: the effect of input spikes period
(the length of repeated spike sequence), and mixing with
non-periodic stochastic spikes.

\begin{figure}
\centering
\includegraphics[width=12cm]{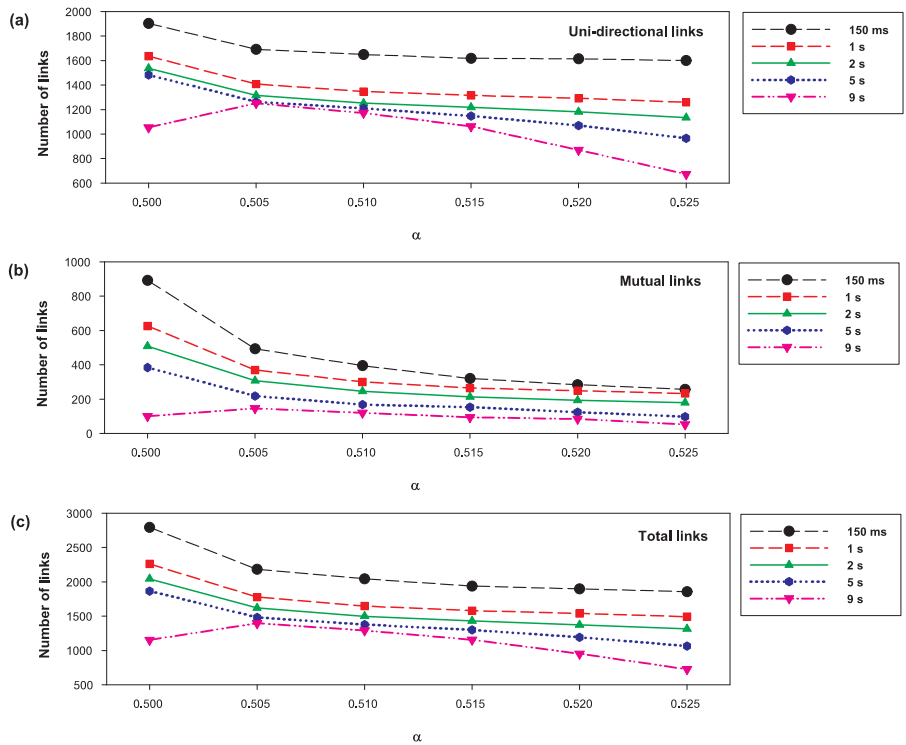}
\caption{\label{fig:PERIOD_B} (color online) The number of surviving
links in networks with input spike trains of different periods:
150ms, 1s, 2s, 5s and 9s. Other parameters are as in the ``basic
configuration".}
\end{figure}

Fig.~\ref{fig:PERIOD_A} shows the effect of input spikes period on
SPs of networks with different asymmetric parameter $\alpha$ of
STDP. We see that the SP does not change when different periods in
the range from 150ms to 9s are used confirming that the particular
SP we found does not depend on the details of the spike sequence,
but the statistical characteristic of a Poisson process.
Fig.~\ref{fig:PERIOD_B} shows the number of surviving links in
networks with input consisting of spike trains of different periods.
For longer periods, the stochastic aspects of the spike trains play
the leading role, and it is more difficult for synapses to survive
under the condition $\alpha\tau_-/\tau_+>1.0$. On the other hand,
for smaller periods, the finite-size effects of the spike trains
play the leading role, and it is easier for synapses to survive. The
neuronal network of \emph{C. elegans} is a very sparse network. By
using a large period, we could also achieve a sparse network that
has a similar number of links as in \emph{C. elegans}, e.g. when
$T_{pattern}=9$s and $\alpha=0.525$.

Fig.~\ref{fig:PERIOD_C} shows the effect of mixing with non-periodic
stochastic spikes on SPs and the final number of links. For the
non-periodic stochastic spikes we also use Poisson processes. The
average firing rate of the mixed spikes that input to neurons are
kept as 50Hz. For a spike train with mixing ratio of 50\% (25\%), we
repeat a spike sequence that is truncated from a 25Hz (37.5Hz)
Poisson spike train, and superimpose a 25Hz (12.5Hz) Poisson spike
train. We find that the particular SP does not change in the case
where we mix the periodic spike trains with non-periodic stochastic
spikes. Because of the stochastic aspect of the spike trains, when a
higher mixing ratio is used, fewer links survive.

\subsection{\label{sec:IVC} The influence of different parameters in STDP}

Next, we analyze the influence of different parameters in STDP, such
as the asymmetrt parameter $\alpha$, the maximum peak synaptic
conductance $g_m$ and the synaptic delay $\tau_d$.

Figs.~\ref{fig:PARAM_A}a and \ref{fig:PARAM_B}a show the influence
of the asymmetry parameter $\alpha$ and the maximum peak synaptic
conductance $g_m$ in STDP rule on the SPs of evolved networks
respectively. From these figures, we find that  the SPs do not
change significantly under parameter variations (see also
figs.~\ref{fig:PARAM_A}b and \ref{fig:PARAM_B}b).  We see that there
are some dependencies between different motifs. For example, the
forward cascade (triad 3) and the FFL (triad 7) are anti-correlated,
while MFFL1 and MFFL2 (triad 9 and 10) as well as the anti-motifs
triads 1 and 2 vary almost identically. As is seen from
fig.~\ref{fig:PARAM_C},  the synaptic delay likewise has little
influence on the SP.

\begin{figure}
\centering
\includegraphics[width=12cm]{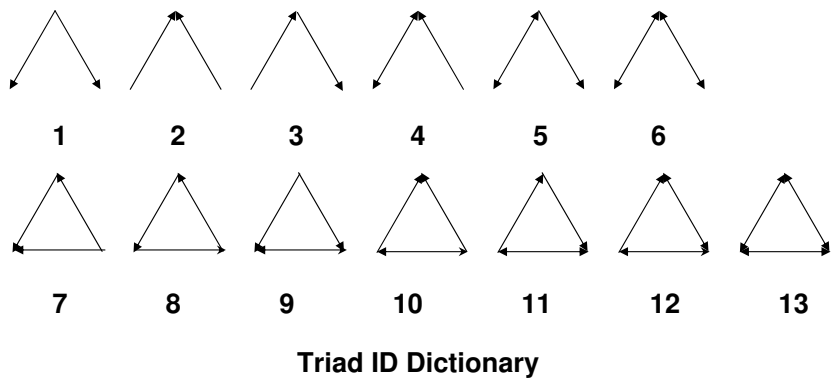}
\caption{\label{fig:Triad} The network structure corresponding to
each triad ID.}
\end{figure}

\begin{figure}
\centering
\includegraphics[width=12cm]{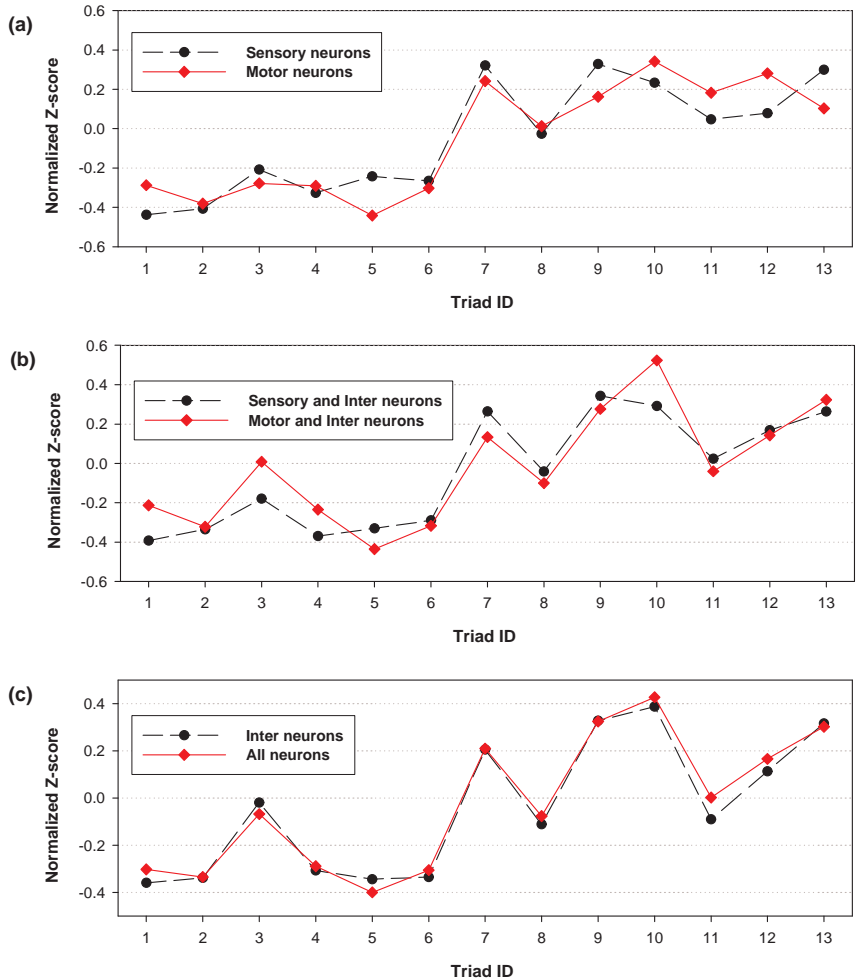}
\caption{\label{fig:CEsp} (color online) The triad significance
profile of different subnetworks in the \emph{C. elegans} neuronal
system. The lines connecting the Z-score values have been added only
as a visualization aid. The structure corresponding to each triad ID
is shown in fig.~\ref{fig:Triad}.}
\end{figure}

\begin{figure}
\centering
\includegraphics[width=12cm]{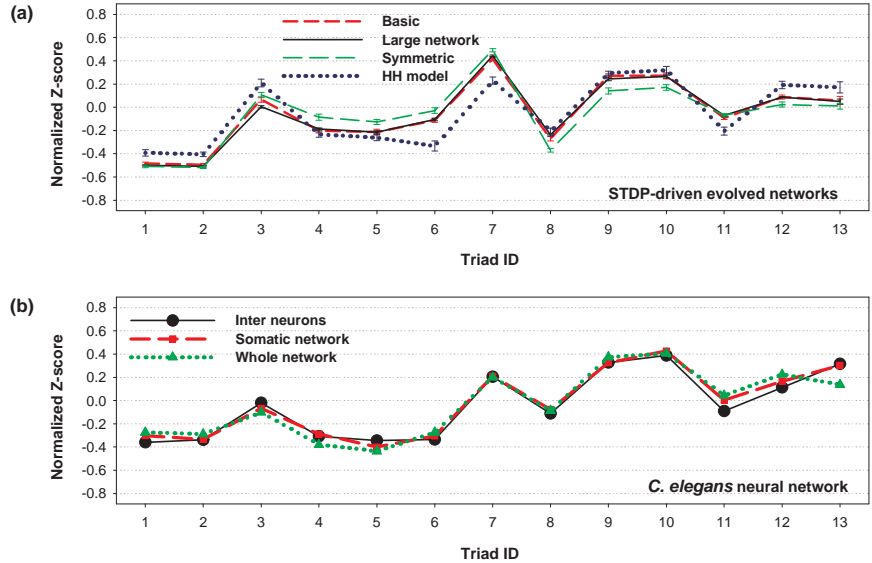}
\caption{\label{fig:STDP_B} (color online) Comparison of SPs for
STDP-driven evolved networks and \emph{C. elegans} neuronal network.
(a) The SPs of four different STDP-driven evolved networks with
different configurations: (i) Basic configuration; (ii) Symmetric
configuration; (iii) HH model configuration; (iv) Large network
configuration. (b) The SPs of \emph{C. elegans} neuronal systems:
(i) the subnetwork of somatic interneurons, (ii) the somatic
network, (iii) the whole neuronal network of the old wiring diagram
in \cite{White}. The structure corresponding to each triad ID is
shown in fig.~\ref{fig:Triad}.}
\end{figure}

\begin{figure}
\centering
\includegraphics[width=12cm]{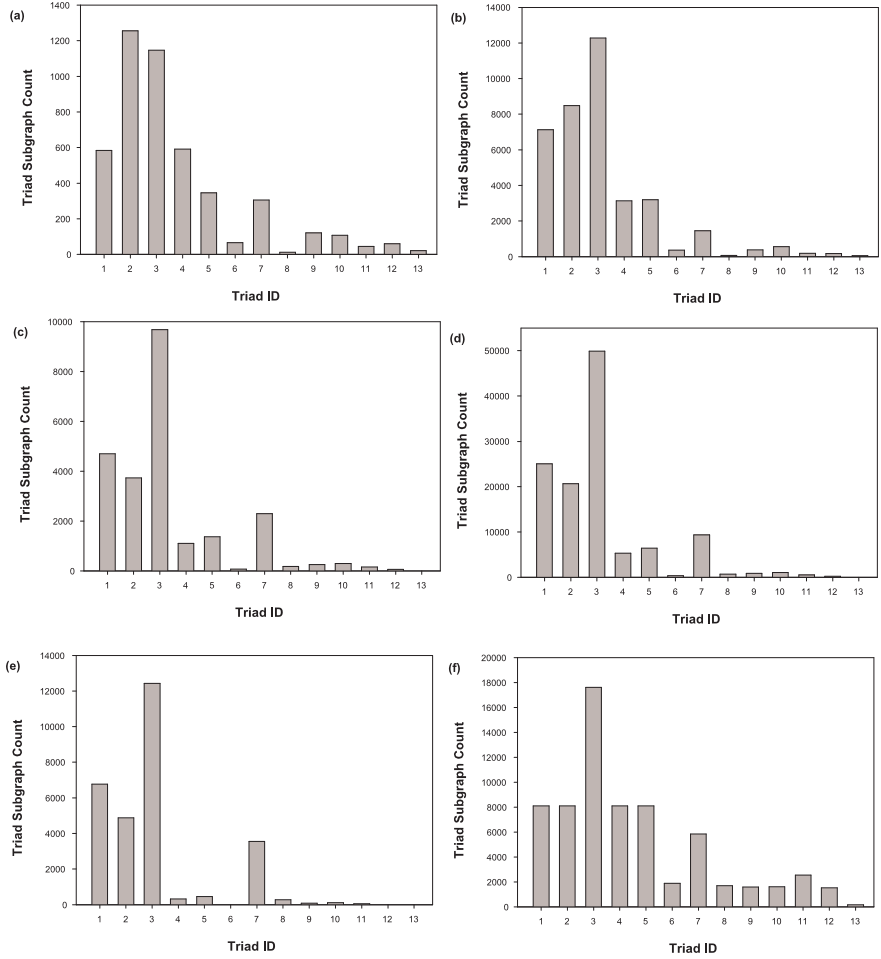}
\caption{\label{fig:STDP_C} Comparison of triad frequency spectra
for STDP-driven evolved networks and \emph{C. elegans} neuronal
network: (a) subnetwork of somatic interneurons in \emph{C.
elegans}, (b) somatic neuronal network in \emph{C. elegans}, and
STDP-driven evolved networks of (c) ``basic configurations", (d)
``large network configurations", (e) ``symmetric configurations",
and (f) ``HH model configurations". The structure corresponding to
each triad ID is shown in fig.~\ref{fig:Triad}.}
\end{figure}

\begin{figure}
\centering
\includegraphics[width=12cm]{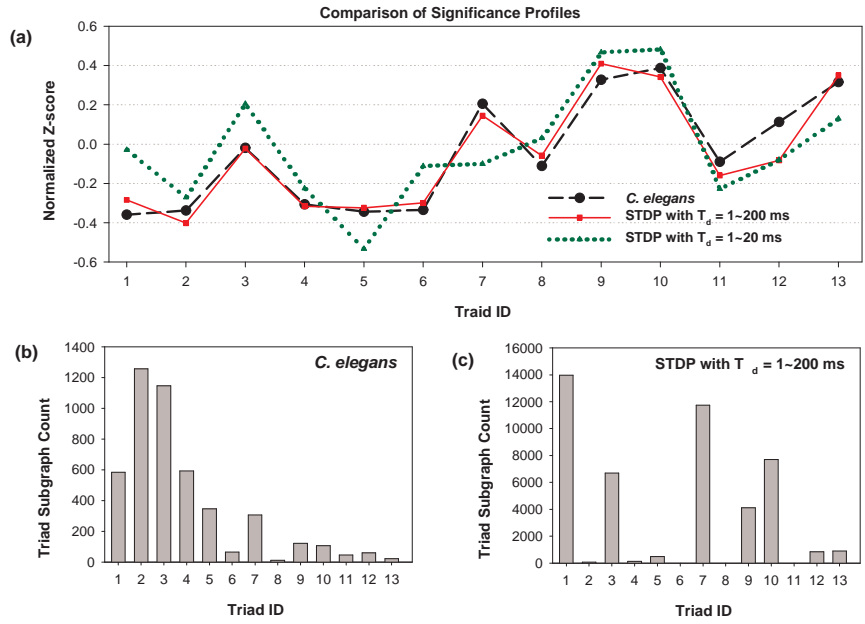}
\caption{\label{fig:CORR} (color online) Comparison of SPs and triad
frequency spectra for the \emph{C. elegans} interneuron subnetwork
and STDP-driven evolved networks with correlated stochastic inputs.
The structure corresponding to each triad ID is shown in
fig.~\ref{fig:Triad}.}
\end{figure}

\begin{figure}
\centering
\includegraphics[width=12cm]{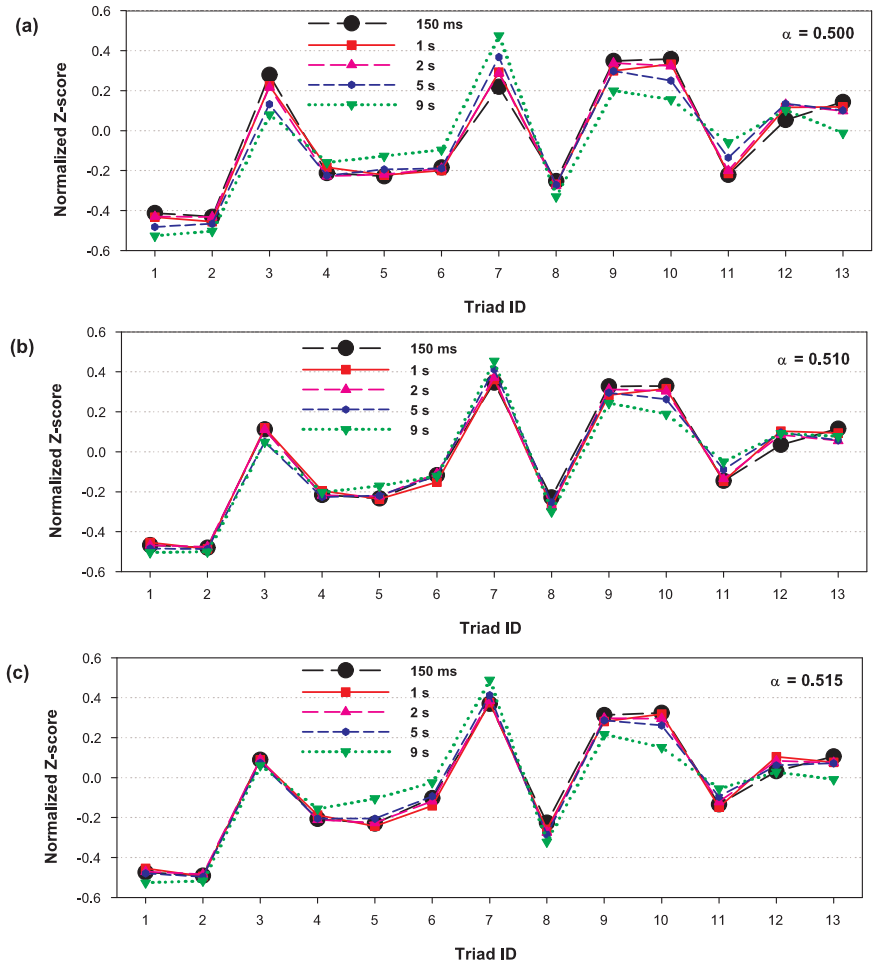}
\caption{\label{fig:PERIOD_A} (color online) The effect of input
spike sequence period $T_{pattern}$ on SPs of STDP-driven evolved
networks with different asymmetric parameter $\alpha$ of STDP rule:
(a) $\alpha=0.500$, (b) $\alpha=0.510$, (c) $\alpha=0.515$.
Different periods $T_{pattern}$ studied are: 150ms, 1s, 2s, 5s and
9s. The structure corresponding to each triad ID is shown in
fig.~\ref{fig:Triad}.}
\end{figure}

\begin{figure}
\centering
\includegraphics[width=12cm]{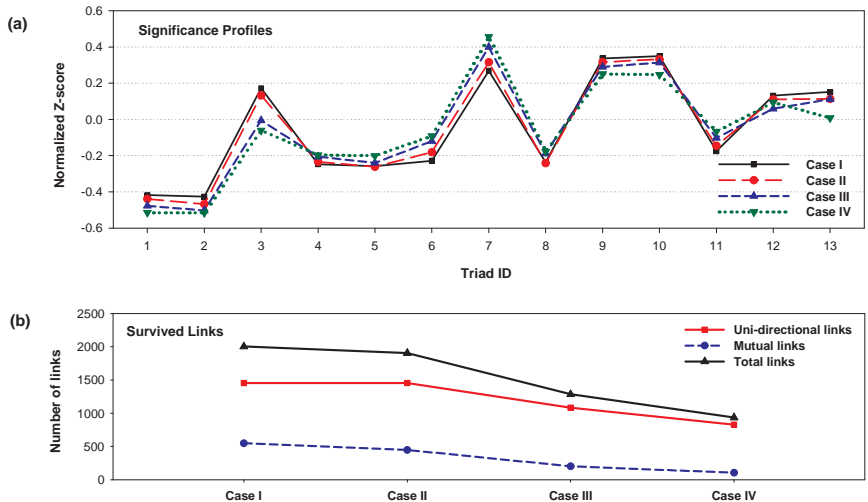}
\caption{\label{fig:PERIOD_C} (color online) The SPs and the number
of surviving links in networks, which receive input of spike trains
mixed with non-periodic Poisson spikes of different average firing
rates: Case I, 13Hz stochastic Poisson processes \& 37Hz periodic
patterns with $\alpha=0.5$; Case II, 25Hz stochastic Poisson
processes and 25Hz periodic patterns with $\alpha=0.5$; Case III,
13Hz stochastic Poisson processes \& 37Hz periodic patterns with
$\alpha=0.55$; and Case IV, 25Hz stochastic Poisson processes \&
25Hz periodic patterns with $\alpha=0.55$. The period of the input
patterns is set to 150ms to save simulation time, while other
parameters are  the same as  in the ``basic configuration". The
structure corresponding to each triad ID is shown in
fig.~\ref{fig:Triad}.}
\end{figure}

\begin{figure}
\centering
\includegraphics[width=12cm]{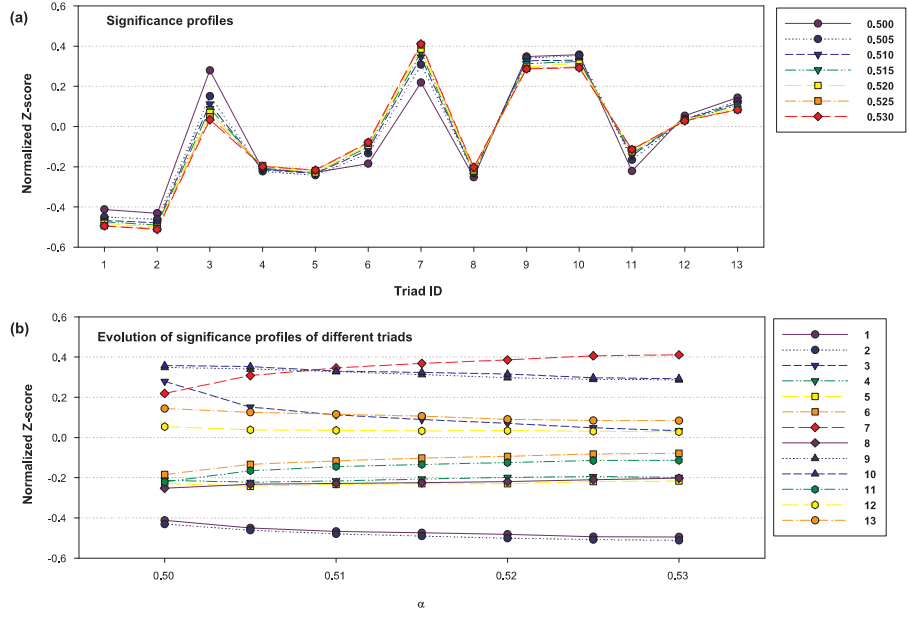}
\caption{\label{fig:PARAM_A} (color online) The influence of the
asymmetry parameter $\alpha$ in STDP rule on the SPs of evolved
networks. (a) The SPs of networks with different $\alpha$. (b) The
evolution of each triad SP along with different $\alpha$. The period
of input patterns is 150ms, while the other parameters are as in the
``basic configuration". The structure corresponding to each triad ID
is shown in fig.~\ref{fig:Triad}.}
\end{figure}

\begin{figure}
\centering
\includegraphics[width=12cm]{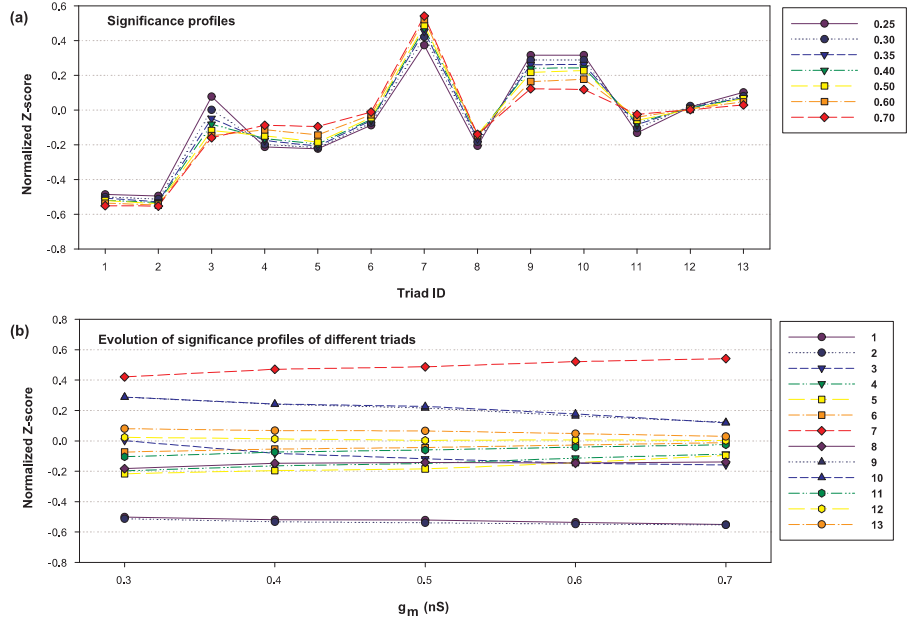}
\caption{\label{fig:PARAM_B} (color online) The influence of the
maximum peak  synaptic conductance $g_m$ in STDP rule on the SPs of
evolved  networks. (a) The SPs of networks with different $g_m$. (b)
The  evolution of each triad SP along with different $g_m$. The
period of input patterns is 150ms, while the remaining parameters
are as in the ``basic configuration". The structure corresponding to
each triad ID is shown in fig.~\ref{fig:Triad}.}
\end{figure}

\begin{figure}
\centering
\includegraphics[width=12cm]{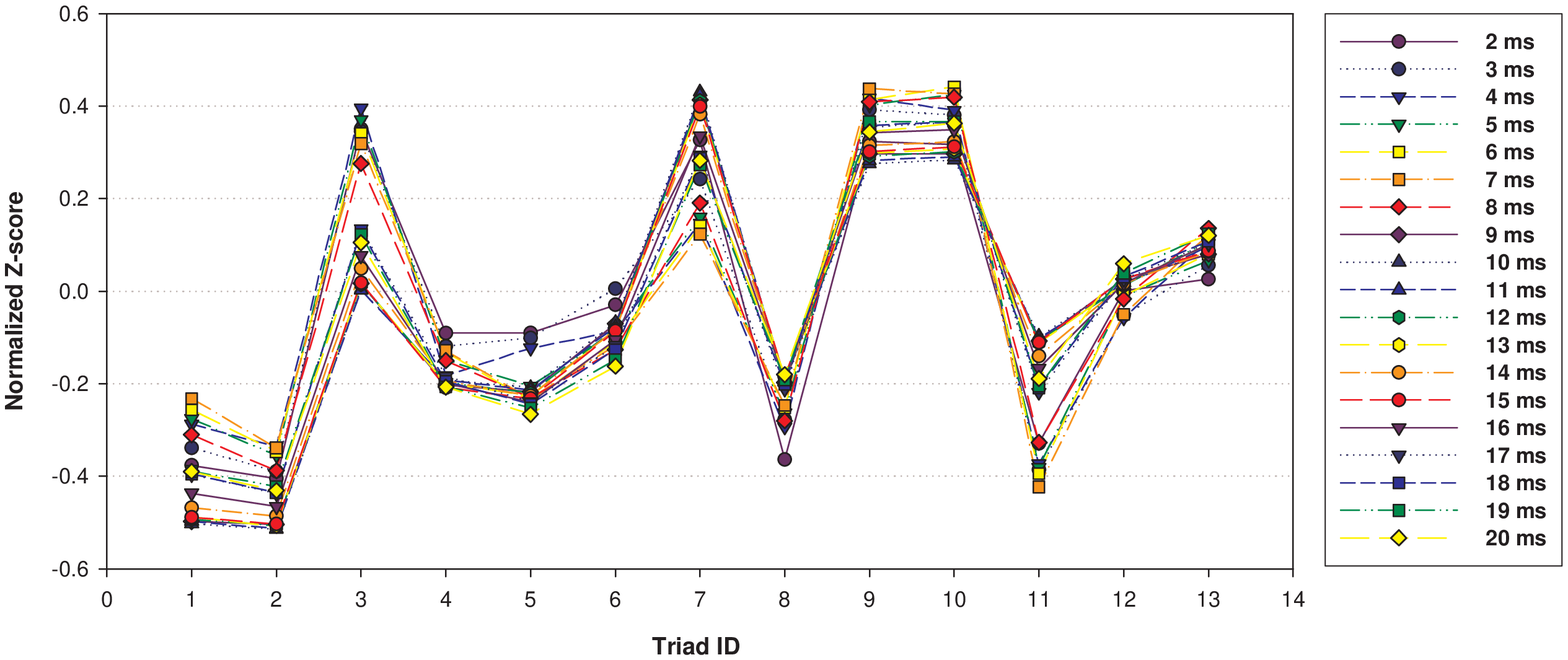}
\caption{\label{fig:PARAM_C} (color online) The influence of
synaptic time delay $\tau_d$ on the SPs of evolved networks. The
period of input  patterns is 150ms, $\alpha=0.55$, while the other
parameters are as in the ``basic configuration". The structure
corresponding to each triad ID is shown in fig.~\ref{fig:Triad}.}
\end{figure}

\section{\label{sec:V}Discussion}
We have studied the effect of the STDP learning rule on the network
evolution systematically. The network is all-to-all connected
initially. Neurons are stimulated with spike trains, which are
(partially) periodic or temporally correlated in a more complicated
way and in line with Poisson statistics. The STDP learning rule
introduces the necessary competition between synapses. As the
network evolves, a stationary distribution of peak synaptic
conductances is achieved, where most synapses become weak enough to
get pruned. In the STDP-driven evolved networks, three important
triads FFL, MFFL1 and MFFL2, which are important for neural
computation and cognition tasks, have been found with much higher
frequency than expected from a randomized network. This implies that
STDP could be a self-organization mechanism that generates these
motifs.

The particular SP we found in STDP-driven evolved networks is quite
robust against parameter variations. The characteristic of SP does
not change essentially no matter what configurations are used, e.g.
different neuron models, asymmetric or symmetric time windows,
different asymmetric ratios $\alpha$, different maximum peak
synaptic conductances $g_m$, different network sizes, different
synaptic delays, different lengths of input sequences, and mixtures
with non-periodic Poisson spike trains. This suggests that we have
found a fundamental characteristic of STDP.

Our simulations mimic the biological case wherein the brain is
densely wired at the time of birth and then most synapses are pruned
in the course of development. To inspect whether STDP may play a
role in the real case, we have compared our simulation results with
the \emph{C. elegans} neuronal system. We have investigated the most
recently published wiring diagram of \emph{C. elegans}, and analyzed
the SPs of different subnetworks in the \emph{C. elegans} neuronal
systems. The SPs of STDP-driven evolved networks are quite similar
to those in the \emph{C. elegans} neuronal network, especially the
subnetwork of interneurons. Besides, the triad frequency spectrum of
STDP-driven evolved network in certain configurations is similar to
that of \emph{C. elegans}. The exact role of input in deciding the
network structure is not yet clear but it seems that some amount of
complexity is needed. The sparsity of the \emph{C. elegans} neuronal
network could be also achieved by the STDP-driven evolved network.
These observations show that the  STDP self-organization mechanism
could be a candidate to generate the local structure of \emph{C.
elegans} neuronal network.

In this study, we only consider STDP synapses and neglect other
mechanisms such as short-term plasticity and gap connections that
are present  in \emph{C. elegans}. Moreover, the neuron models and
parameter setting may also be different from \emph{C. elegans}.
These factors all could influence the results. For example, gap
connections are bi-directional in \emph{C. elegans}, and could
influence the dynamics of neurons. As the STDP learning rule depends
on spiking activity, the existence of gap connections could
influence the structure of an STDP-driven evolved network. The few
observed differences between the simulated networks and the \emph{C.
elegans} neuronal network may be due to one of these reasons. In any
case, the similarities between the two networks are striking
considering the fact that such other complications have not yet been
included.

The method of motif analysis we used is based on the algorithm of
Milo \emph{et al.} \cite{Milo:superfamily}, which is widely used in
different scenarios. Randomization is performed by rewiring
connections. It features include: It keeps the number of incoming
edges, outgoing edges and mutual edges of each node fixed. Motifs
are counted here on the basis of the number of vertices they
involve. So, for instance, triad 1 is a motif at the same level as,
for instance, 7 because they both involve 3 vertices. In particular,
1 is not counted as a submotif of 7 because 7 contains an edge that
is absent in 1. Conceptually, a problem with this approach is that
such motif counts are not statistically independent. In particular,
keeping the counts for certain triads fixed may not be a good null
model in statistical terms.  A more principled approach is developed
in \cite{Kahle,Olbrich}, based on correlations between edges, in
such a manner that the motifs are arranged in terms of the number of
edges they contain, as opposed to the number of vertices involved.
We cannot go into the details, but the underlying rationale is that
a network is characterized by the specific presence or absence of
edges for some fixed set of vertices, and null models therefore
should be given in terms of correlations up to some given number of
edges. This, however, means that for instance, triad 1 becomes a
submotif of 7 so that each occurrence of 7 also counts as an
occurrence of 1. So far, however, that approach has been developed
only for undirected graphs. In any case, when we base the comparison
of SPs between our STDP-driven evolved artificial network and the
\emph{C. elegans} neuronal network on that method, the similarities
become much weaker. This may come about because by double counting,
the SP of triads will be perturbed if there are variations in the
SPs of more complex triads. These triads may be independent
functional units when we start with a fully connected network and
prune synapses according to the STDP learning rule, as opposed to
beginning with a completely unconnected set of vertices and
incrementally adding edges. It needs future work to explore this
issue further.

The study suggests that the structure of a neuronal network is
mostly determined by a correlation based learning mechanism. In
particular, this seems to be based on self-organization on the basis
of a local rule  as opposed to a hard wired network scenario. The
dynamical steady state reached also depends on the nature of the
input supplied to neurons. This work identifies the important
factors that determine the structure of the network. The conclusions
might be useful in the choice of the neuronal substrate in large
scale neural network simulations of the brain.

\section*{Acknowledgments}
The authors thank Nils Bertschinger, Bernhard Englitz, Thomas Kahle
and Eckehard Olbrich for discussions. KMK acknowledges a research
grant from Department of Science and Technology (DST), India. JJ
acknowledges support from the Volkswagen Foundation.


\newpage 

\end{document}